\documentclass[prl,twocolumn,showpacs,superscriptaddress]{revtex4-1}

\usepackage{amsmath,amssymb, amsfonts}
\usepackage{bm}
\usepackage[usenames,dvipsnames]{color}
\usepackage{graphicx}

\def\tr{{\rm Tr} \, }
\def\pmx{\begin{pmatrix}}
\def\emx{\end{pmatrix}}

\def\be{\begin{eqnarray}}
\def\ee{\end{eqnarray}}
\def\bee{\begin{eqnarray*}}
\def\eee{\end{eqnarray*}}

\def\<{\langle}
\def\>{\rangle}
\def\ot{\otimes}
\def\tr{{\rm Tr} \, }
\def\trp{{\rm Tr} }
\def\kb{ \> \< }

\def\half{\tfrac{1}{2}}
\def\wtd{\widetilde}
\def\wh{\widehat}

\def\bra{\langle}
\def\ket{\rangle}
\def\ot{\otimes}
 \newcommand{\proj}[1]{ | #1 \kb  #1|}
\def\tr{{\rm Tr} \, }
\def\trp{{\rm Tr} }
\def\kb{ \ket \bra }

\def\ket{\>}
\def \NBodyDM{\ensuremath{\Lambda}}

\newcommand{\myeqref}[1]{{Eq.\eqref{#1}}}
\newcommand{\eqpunct}[1]{#1}

\newcommand{\expt}[3]{\ensuremath{ \<#1| {#2} |#3\> }}

\def\latt{{\rm latt}}
\def\sing{{\rm sing}}
\def\doub{{\rm doub}}
\def\ferm{{\rm ferm}}

\def \LogZ {\wtd{Z}}
\def \LogX {\wtd{X}}
\newcommand{\XOP}[1]{\ensuremath{X_{#1}}}

\newcommand{\ZOP}[1]{\ensuremath{Z_{#1}}}
\renewcommand{\forall}{\text{ for all }}

\begin{document}

\title{Quantum codes give counterexamples to the unique pre-image conjecture of the $N$-representability problem}

\author{Samuel A. Ocko}%
\affiliation{Department of Physics, Massachusetts Institute of
  Technology, Cambridge, Massachusetts, USA}%
\author{Xie Chen}%
\affiliation{Department of Physics, Massachusetts Institute of
  Technology, Cambridge, Massachusetts, USA}%
\author{Bei Zeng}%
\affiliation{Department of Mathematics and Statistics, University of Guelph, Guelph, Ontario, Canada}%
\affiliation{Institute for Quantum Computing, University of Waterloo, Waterloo, Ontario, Canada}%
\author{Beni Yoshida}%
\affiliation{Department of Physics, Massachusetts Institute of
  Technology, Cambridge, Massachusetts, USA}%
\author{Zhengfeng Ji}%
\affiliation{Perimeter Institute for Theoretical Physics, Waterloo,
  Ontario, Canada}%
\author{Mary Beth Ruskai}%
\affiliation{Department of Electrical and Computer Engineering, Tufts University, Medford,
  Massachusetts, USA}%
\author{Isaac L. Chuang}%
\affiliation{Department of Physics, Massachusetts Institute of
  Technology, Cambridge, Massachusetts, USA}%

\date{\today}

\begin{abstract}
It is well known that the ground state energy of many-particle Hamiltonians involving only $2$-body interactions can be obtained using constrained optimizations over density matrices which arise from reducing an $N$-particle state. 
While determining which $2$-particle density matrices are ``$N$-representable'' is a computationally hard problem, all known extreme $N$-representable $2$-particle reduced density matrices arise from a unique $N$-particle pre-image, satisfying a conjecture established in 1972.
We present explicit counterexamples to this conjecture through giving Hamiltonians with $2$-body interactions which have degenerate ground states that cannot be distinguished by any $2$-body operator.
We relate the existence of such counterexamples to quantum error correction codes and topologically ordered spin systems.
\end{abstract}

\maketitle


For all known systems of identical particles, which have Hamiltonians
 restricted to symmetric (bosonic) or anti-symmetric (fermionic)
states, the Hamiltonians contain at most two-body interactions.
Therefore, for many purposes such as energy calculations, an $N$-body state can be replaced by its
$2$-particle reduced density matrix (RDM). In doing so, one might hope to reduce complex $N$-particle variational calculations with simpler $2$-particle ones. Early efforts gave absurdly low energies until it was realized that it was necessary to restrict energy minimization to $2$-particle density
matrices which, in fact, come from the reduction of an $N$-particle state
of the appropriate symmetry (its {\em pre-image}). Characterizing
these $2$-particle RDMs is a fascinating question known as the
$N$-representability problem \cite{Coleman63,Garood64a,Coleman00}.

In the 1960's, the $N$-representability problem was solved for $1$-particle RDMs \cite{Coleman63}. However, finding a solution for $2$-particle RDMs is so challenging
that most of those who tried concluded that it was intractable. This
intuition was recently validated with a quantum information theoretic
proof that $N$-representability for the $2$-particle RDM belongs to the
complexity class called QMA complete \cite{Liu07}, i.e., the worst
cases would be hard even with a quantum computer.

Surprisingly, this coincided with a revival of interest in the
$N$-representability problem from several directions. A number of
groups have obtained good approximations to the ground state energy in
special situations \cite{Mazziotti01a, *Cances06a,*Nakata08a,*Mazziotti08a,*VanAggelen10}.
New eigenvalue bounds for the $1$-particle RDM have been found for pure
$N$-fermion states \cite{Altunbulak08a}. For both fermions and
bosons the first improvements on expectation value bounds since 1965
were obtained in \cite{Shenvi06a}. Moreover, the map from an $N$-particle state to a $m$-particle RDM is a special type of quantum channel, which found an important application involving Renyi entropy in \cite{Grudka09a}.

A widely held property about the convex set of $N$-representable $2$-particle RDMs is that every extreme point has a unique $N$-particle pre-image. \emph{Extreme $N$-representable RDMs} are fundamental; every $N$-representable $2$-particle RDM is a weighted average of extreme points. Because energy is a linear function of the RDM, its minimum must lie on the set of extreme points. The unique pre-image
property holds for the best known extreme points, which come from generalizations of BCS states \cite{Erdahl72a,Coleman00}. It is also true for the few
other known extreme RDMs; similar observations \cite{Verstraete06} have been made for RDMs of translationally symmetric spin lattice systems. In 1972, Erdahl \cite[Section 6] {Erdahl72a} formally
conjectured that all extreme RDMs have a unique pre-image. This has been  believed to hold since then. 

Erdahl's conjecture has been proven for $m$-particle extreme RDMs when $2m >N$ \cite{Erdahl72a}. Moreover, if the conjecture were false,
there would exist an unusual $2$-body Hamiltonian, whose
ground state degeneracy is ``blind'' to, i.e.
undetectable by, $2$-body operators. All ground states of such a
``$2$-blind'' Hamiltonian would have the same $2$-particle RDM, and thus the degeneracy cannot be broken without at least a $3$-body interaction; $2$-body perturbations would only shift the energy.

In this paper, we give explicit counterexamples to Erdahl's conjecture. To do so, we first exhibit a class of $2$-blind \emph{spin lattice} Hamiltonians, whose ground states are quantum error correction codes. Extended to fermions, these examples provide extreme $N$-representable  $2$-particle RDMs with multiple pre-images, which
are thus the desired counterexamples. We then directly relate the
general conditions for quantum error correction to the existence of
such counterexamples. Our results imply that extreme
$N$-representable RDMs can be very different than those known previously. In addition, the Hamiltonians we use \cite{Bacon06a}
play a pivotal role in the study of topological quantum error
correction \cite{Kitaev02, Bombin:2006}.


\smallskip \noindent{\bf $N$-representability and the unique pre-image conjecture:} We begin with a brief
description of the $N$-representability problem and its generalization
beyond fermionic symmetries. For fermions, a symmetric $2$-body Hamiltonian
$H_N $ acts on anti-symmetric states $| \psi^- \ket $ for which the
$2$-particle RDM is $\rho^{-}_{12} = \trp_{3 \ldots N} \proj{\psi^-} $. A $2$-particle RDM $\rho^{-}_{12}$ is called \emph{$N$-representable} if it has a \emph{pre-image} $\Lambda$, i.e.  $\rho_{12}^- = \trp_{3 \ldots N} \Lambda $, where $\Lambda = \sum{p_{\psi^-}\proj{\psi^-}}$ is an $N$-fermion state. The critical interplay between the one($T_j$) and two-body$(V_{jk})$ terms of $H_N$ is captured by
\be
     H =  H_N - E_0 = \sum_j T_j +  \sum_{j < k} V_{jk} ~ - E_0 
         = \sum_{j < k } \wh{H}_{jk}^N  \eqpunct{,} \quad
\ee
where $E_0$ is the ground state energy of $H_N$, and $\wh{H}_{jk}^N \equiv
V_{jk} + \tfrac{1}{N-1} (T_j + T_k) - \tbinom{N}{2}^{-1} E_0 $ is known as the
reduced Hamiltonian. One can verify the energy of $\Lambda$ is determined by its $2$-particle
RDM $\rho^{-}_{12}$ by
\be 
  \tbinom{N}{2}  \tr  \wh{H}_{12}^N \,  \rho^{-}_{12} = \tr H \NBodyDM  \geq 0 \eqpunct{,} \label{dual}
\ee  
with equality if and only if $\NBodyDM$ is a ground state of $H_N$. 

Although $H$ is positive semidefinite by construction,
$\wh{H}_{12}^N $ is \emph{not} positive semidefinite in general;
however, it acts as if it were on the set of $N$-representable RDMs.
Thus it acts as a ``witness'' for $N$-representability, a special case of a general duality concept known as the polar cone of a convex set.
 
\begin{figure}[htbp]
\centerline{\includegraphics[width=7cm]{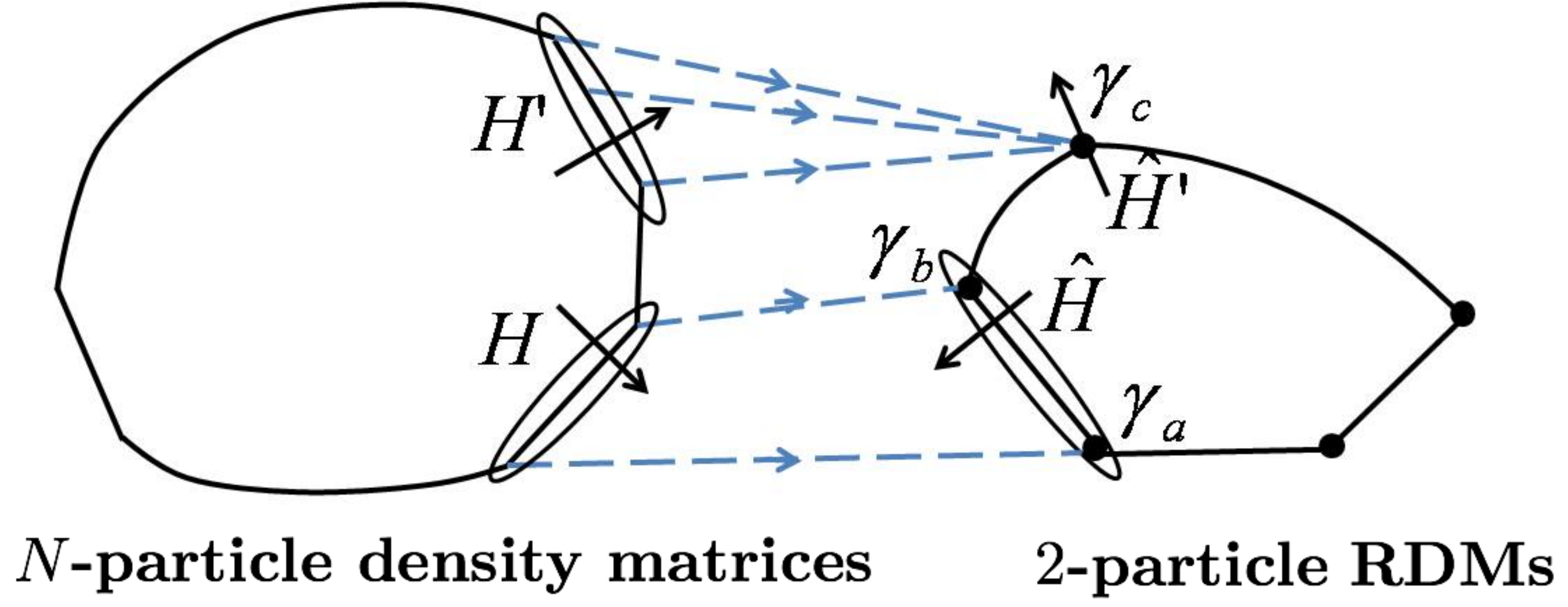}}
\caption{Mapping of $N$-particle density matrices to $N$-representable
 $2$-particle RDMs. $\gamma_a, \gamma_b$ are extreme  $2$-particle RDMs with unique
 pre-images. $\gamma_c$ is an extreme $2$-body RDM with multiple
 pre-images.}
\label{fig:nrep}
\end{figure}

$N$-representable $2$-particle RDMs form a convex set, as the average of two is also $N$-representable. The set of $N$-representable RDMs, like any convex set, is characterized by its extreme points, which are not the average of any two points in the set (Fig.~\ref{fig:nrep}). Erdahl \cite[Section~3]{Erdahl72a} showed that in finite dimensions every extreme $N$-representable RDM
$ \gamma_{12}$ is also {\em  exposed} in the sense that there is some
Hamiltonian $\wh{H}_{12}^N$ for which $\gamma_{12}$ is the unique
lowest energy $N$-representable RDM. Every extreme point thus
corresponds to the ground state eigenspace of at least one two-body Hamiltonian
$H_N$.

When the ground state of $H_N$ is non-degenerate, it is the unique
pre-image of its $2$-particle RDM. A degenerate ground state eigenspace defines a
convex subset of the $N$-representable $2$-particle RDMs, which typically
corresponds to a flat (exposed) region on the boundary. In exceptional cases, this region is a single extreme point with multiple pre-images; this happens when the Hamiltonian is $2$-blind, meaning that all degenerate ground states have the same $2$-body RDM.

It is useful to extend the concept of $N$-representability to
the complete absence of symmetry. This leads to the closely
related {\em quantum marginal} problem which asks if there is
a pre-image $\Lambda = \sum{p_{\psi}\proj{\psi}}$ consistent with
the reduction to $\{\rho_{jk}\} \equiv ( \rho_{12} , \rho_{13} , \ldots, \rho_{1N}, \rho_{23}, \ldots, \rho_{N-1 \text{,}N})$, where $\{\rho_{jk}\}$ is expressed in vector form to emphasize that those
which are consistent form a convex set. The reduced spin lattice Hamiltonian $\{ \wh{H}_{jk} \}$ is also written in vector form, and \myeqref{dual} becomes $\sum_{j < k} \tr \wh{H}_{jk} \rho_{jk} $.
       
Erdahl's conjecture is equivalent to the statement that there is no $2$-blind \emph{fermionic} Hamiltonian. We first present a $2$-blind \emph{spin lattice}
Hamiltonian which gives an extreme quantum marginal $\{\rho_{jk}\}$ with multiple pre-images. We then explain how to extend this to fermions to disprove Erdhal's conjecture.

\smallskip \noindent{\bf Lattice example:} We consider the Hamiltonian
for the two-dimensional quantum compass model used in condensed matter
physics \cite{Kugel82a,*Dorier05a}. It has a doubly degenerate ground state
eigenspace, known as the Bacon-Shor code \cite{Bacon06a} in quantum
information theory, for which the  $2$-particle RDMs $\{\rho_{jk}\}$ are independent
of the choice of eigenstate.

Let $X_{jk}$ and $Z_{jk}$ denote the Pauli operators $\sigma_x $ and
$\sigma_z$, respectively, acting on the site $(j,k)$ in a square $n
\times n$ spin lattice, and define
\be   \label{bacon}
     H_{n^2}\equiv   - \sum_{jk} \big(  J_x X_{j,k} X_{j+1, k }  + J_z Z_{j,k} Z_{j,k+1}   \big) \eqpunct{,}
\ee
where $J_x, J_z > 0$, and subscript addition is mod $n$, corresponding to cyclic boundary conditions.

For $n=3$, define the even parity columns
 \setlength\abovedisplayskip{10pt}
\bee
    v_0 = \pmx 0 \\ 0 \\ 0 \emx \eqpunct{,} \quad   v_1 = \pmx 0 \\ 1 \\ 1 \emx \eqpunct{,} \quad   
       v_2 = \pmx 1 \\ 0 \\ 1 \emx \eqpunct{,} \quad  v_3 = \pmx 1 \\ 1 \\ 0 \emx \eqpunct{.}
\eee
The odd parity columns are obtained by flipping all spins $0
\leftrightarrow 1$. Symmetry considerations can be used to show that
$H_{n^2}$ is block diagonal, including two blocks spanned by states
where all columns have even parity or all columns have odd parity. 

It can be shown that the lowest energy eigenvalues of these blocks are unique
and those of other blocks strictly larger
\cite{Bacon06a,Doucot05a}. (In fact it is not hard to verify that $E_0 \leq -9J_Z$ on these blocks and that for $J_Z \gg J_x$ the Hamiltonian
is $\gtrsim -5J_Z$ on the other blocks.) Therefore, this Hamiltonian has a
pair of ground states which can be distinguished by even and odd
parity.  Let

\begin{flalign*}
|A_1 \>  & =   \half   \sum_{j =0}^{3}   (v_j , v_j, v_j ) \eqpunct{,  } \text{  }  |A_3 \>    =   \tfrac{1}{\sqrt{24} } \sum_{i \neq j \neq k \neq i}  (v_i, v_j, v_k)  \eqpunct{,}   \\
|A_2 \> &=   \tfrac{1}{6} \sum_{j \neq k}  \big[ (v_j, v_k, v_k) + (v_k, v_j, v_k)  + (v_k, v_k, v_j)  \big]  \eqpunct{.}
\end{flalign*}

The  even parity ground state of \myeqref{bacon} is
\be
     |C_0 \> = a _1 |A_1 \>  + a_2 |A_2 \>  + a_3  |A_3 \> \eqpunct{,}
\ee
and the odd parity ground state $|C_1 \>$ can be obtained by flipping all spins.

To verify that all ground states $u |C_0 \> + v|C_1 \>$ have the same
 $2$-particle RDMs, it is both necessary and sufficient to show that for
all $2$-body operators $B$,
\be
   \expt{C_p}{B}{C_q} = 
   \delta_{pq} b 
\label{indistinguishabilitycriteria} 
\ee
for some constant $b$. To prove that $\expt{C_p}{B}{C_p} = b \forall p$, we
introduce the parity-conversion operators $\LogX_{j} = \prod_i{\XOP{j, i}}$, which change the parity of the ground states; $\LogX_{j} |{C_p}\> = |{C_{1-p}}\>
\forall j$. Given any $2$-body operator $B$ which acts on sites $(j_1,
k_1)$ and $(j_2, k_2)$, an $\LogX_{j}$ may be chosen which does not
affect the same sites as $B$; $j \neq j_1, j_2$. Therefore:
\be
 \expt{C_0}{B}{C_0} = \expt{C_0}{\LogX_j B \LogX_j}{C_0} = \expt{C_1}{B}{C_1} = b\eqpunct{.}
\ee
One can similarly show $\expt{C_0}{B}{C_1} = 0$ using the phase-flipping operator $\LogZ_{k} =
\prod_i{\ZOP{i, k}}$, $\LogZ_{k}  |{C_p}\>  = (-1)^p  |{C_p}\>$.

Therefore, because $H_{n^{2}}$ is a $2$-body Hamiltonian
where all degenerate ground states have the same $2$-particle RDMs, it follows that $H_{n^2}$ is $2$-blind and $\wh{H}_{n^2}$ gives an exposed, and therefore extreme $\{\rho_{jk}\}$ with no unique pre-image. This fact is already surprising and interesting. We show next that this can be mapped to fermions to give a counterexample to Erdahl's conjecture.

\smallskip \noindent{\bf From spin lattices to fermions:}  
To extend our example to fermions,
we introduce orthonormal site label functions $\{ f_1, f_2, \ldots,  f_N \}$  
which one can be regard as
 the spatial components of a full electron wave function. 
 Defining $a_{j, s_j}^\dag $ to be the
 fermion creation operator
 $a_{j, s_j}^\dag | \Omega \ket = | f_j s_j \ket$ with $| \Omega \ket $
 the vacuum, we map each lattice basis state to a Slater determinant having one fermion per site:
\be
         V: | s_1 \ket \ot \ldots \ot   | s_N \ket  \mapsto  
         a_{1 , s_1}^\dag \ldots   a_{N , s_N}^\dag | \Omega \ket \label{StateMapping} \eqpunct{.}
\ee     
The map $V$ can be extended by linearity to arbitrary lattice
states $V: |\psi \ket \mapsto |\psi^- \ket$ so that $V^\dag V = I_{\latt} $
and $V V^\dag$ projects onto the anti-symmetric subspace
 ${\cal H}_{\sing}^{-} $ with 
half-filled (singly occupied) spatial orbitals.
We also define maps
$V_{jk} : |s_j s_k \ket \mapsto a_{j, s_j}^\dag a_{k, s_k}^\dag | \Omega \ket $
 restricted to sites $j,k$; one can then verify that $\rho_{12}^{-}  =  \trp_{3 \ldots N} \proj{\psi^{-}}$
 can also be directly obtained from $\{\rho_{jk} \}$:
 \begin{equation*}
\{\rho_{jk}\}  \mapsto \tbinom{N}{2}^{-1}  \cdot   \! \sum_{1 \leq j <k \leq N}{V_{jk} \rho_{jk} V_{jk}^{\dag} } =  \rho_{12}^- \eqpunct{.}
   \end{equation*}
 Because $\rho_{jk} = V_{jk}^\dag \rho_{12}^- V_{jk}$, we can
 also recover $\{\rho_{jk}\}  $ from $\rho_{12}^{-}$ . Since $V$ acts like a unitary map from the lattice space to ${\cal H}_{\sing}^{-} $, we can obtain the full correspondence: 
   \be     \label{diagram}
 \begin{array}{ccc}
         |\psi \ket   &  \longleftrightarrow  &  |\psi^{-} \ket  \\
            \downarrow & & \downarrow \\
            \{ \rho_{jk} \}   &  \longleftrightarrow  & \rho_{12}^{-}
            \end{array}
        \ee
 Since $\leftrightarrow$ denotes one-to-one mappings, it
 follows immediately that $\rho_{12}^{-}$ is extreme with multiple pre-images 
if and only if $ \{ \rho_{jk} \} $ is extreme with corresponding pre-images. 
Applying this procedure to the Bacon-Shor code gives an extreme fermionic $2$-particle RDM with multiple pre-images, disproving Erdahl's conjecture. Moreover, because every extreme $N$-representable $2$-particle RDM is exposed, we know that there exists a ``$2$-blind'' fermionic Hamiltonian which exposes this extreme point.

To construct an exposing Hamiltonian, 
we first define a map on traceless one-body spin operators $W_{j}$ as:
 \be
W_j = \sum_{st} w_{st}|s_{j} \ket \bra{t_j}| \mapsto W_j^{\ferm} = 
  \sum_{st} w_{st} a^{\dag}_{j,s_j} a_{j,t_j}\eqpunct{.} \quad \label{OperatorMapping}  
\ee
Since any lattice Hamiltonian can be written as a linear combination of
tensor products of one-body operators, \myeqref{OperatorMapping} 
maps any lattice Hamiltonian to a fermionic $H_{\ferm}$, acting on the 
 \emph{full} anti-symmetric subspace. $H_{\ferm}$ conserves particle number at each site and acts on the invariant subspace ${\cal H}_{\sing}^{-} $ in the same way that $H_{\latt}$ acts on the lattice space: 
 $$ H_{\latt} \mapsto  H_{\ferm} = \pmx H_{\sing}& 0 \\  
                   0 &  H_{\doub} \emx \eqpunct{, } \phantom{aa} H_{\sing} = V H_{\latt} V^\dag \eqpunct{.} $$
    To ensure that the ground state of $H_{\ferm}$ is that
of $H_{\sing}$, we add a penalty term $\sum_{j}  U_j \left( n_{j}-1 \right)^2$ with sufficiently large
 $U_j$, where $n_{j} = a_{j,\uparrow}^\dag a_{j,\uparrow} + a_{j,\downarrow}^\dag a_{j,\downarrow}$.
This gives an $n^2$-parameter family of exposing Hamiltonians.

These procedures also work for bosonic systems.

\smallskip \noindent{\bf Quantum error correction codes:} The
counterexample given above is a special case of a much more general
connection between RDMs with multiple pre-images, $m$-blind
Hamiltonians, and quantum error correction codes. In quantum coding theory \cite{Nielsen00a}, a quantum state is encoded into a subspace of a larger system in a way such that errors can be identified and corrected without disturbing the encoded state. This subspace is spanned by an orthonormal basis of codewords $|C_p\>$, and a necessary and sufficient condition for a quantum code to be able to correct a set of single particle errors ${\cal E} = \{ E_m \}$ is that:
\be
\< C_p |E^\dagger_\ell  E_m |C_q \> = \delta_{pq} Q_{\ell m} \eqpunct{.}
\,
\ee   
${\cal E}$ contains the operators $\{F_{1,a}, \ldots F_{N,a} \} \forall a$, where $F_{j,0}, F_{j,1}
\ldots F_{j, d^2-1}$ is a basis of for one-particle operators on site $j$. Therefore, since $E^{\dag}_{\ell} E_{m} = F_{j,a}^{\dag}F_{k, a'}$ forms a basis for the set of two-body operators, the criteria for a code to be able to correct all single-particle errors $\< C_p |E^\dagger_\ell  E_m |C_q \> = \delta_{jk} Q_{\ell m} $, is \emph{exactly} the criteria of
\myeqref{indistinguishabilitycriteria} for all states in the code
space $\{|C_p \ket\}$ to have the same set of $2$-particle RDMs. 

This set of $2$-particle RDMs will be extreme if and only if the code space is the ground space of some Hamiltonian with at most $2$-body interactions. The Bacon-Shor code has this property and yields extreme points with multiple pre-images. However, most quantum codes, including stabilizer codes and non-stabilizer CWS codes \cite{Nielsen00a, Cross09a}, do not have this property and simply yield interior or boundary points with multiple pre-images.

Erdahl's general conjecture was for $m$-particle RDMs, with $m \geq 2$. While counterexamples for $m > 2$ can come from the Bacon-Shor code defined on larger lattices, they also come from $m$-blind Hamiltonians whose ground states define a quantum code that can correct any $\lfloor \frac{m}{2} \rfloor$-particle errors. Topological quantum codes can have this property; for example, Kitaev's toric code \cite{Kitaev02} is a $4$-blind Hamiltonian which gives an extreme $N$-representable $4$-particle RDM with multiple pre-images. Other topological quantum codes exhibit the same properties \cite{Bombin:2006, MacKay04a}. Indeed, similar relationships between topological quantum codes and RDMs have been observed in \cite{Kitaev20032, Bravyi:2010fk}.


\smallskip \noindent{\bf Extensions and open issues:} 
The fermionic extreme points constructed here, which come from $N$-particle states with half-filled orbitals, are quite different from those one encounters for atomic and molecular systems, and also differ from the best known extreme points which come from generalizations of BCS states \cite{Erdahl72a,Coleman00}. The critical issue is not whether the states described here -- or their associated Hamiltonians -- arise in practical applications. Our results demonstrate that the class of extreme points is much larger and more complex than previously believed. From the standpoint of quantum chemistry, the challenge is to characterize a class of extreme points which will lead to useful new computational algorithms.

The $2$-blind fermionic Hamiltonians we used to disprove Erdahl's conjecture are quite different from fermionic Hamiltonians that physicists usually encounter, which have two-body potential terms as well as one-body terms having the form of a Laplacian. This leads to a question of fundamental importance in developing physically realizable quantum codes; can Hamiltonians with physically reasonable Laplacian and local potential terms, including realistic spin and magnetic interactions, be $2$-blind?

Some of these counterexamples are closely related to topologically ordered spin systems. Stabilizer topological codes are counterexamples for $m >2$, and subsystem \cite{ Kribs05a} topological codes \cite{Bombin10a} are candidates of counterexamples for $m=2$. 
Further work along these directions will undoubtedly continue to forge
new connections between quantum information, condensed matter physics,
and quantum chemistry.

\begin{acknowledgments}  
This work was begun when BZ, ZJ, and MBR were visiting the Institute for Mathematical Sciences, National University of Singapore in 2010. 
SAO is supported by NSF Grant No. DGE-0801525, {\em IGERT: Interdisciplinary Quantum Information Science and Engineering}.
XC acknowledges support by the NSF CUA. 
ZJ acknowledges support by Industry Canada and the Ministry of Research \& Innovation. 
BZ is supported by NSERC and CIFAR.
MBR and ILC are supported by the NSF. 
\end{acknowledgments}

%

\end{document}